\documentclass[twocolumn,showpacs,superscriptaddress,preprintnumbers,amsmath,amssymb,longbibliography,prl]{revtex4-1}
\usepackage{graphicx}
\usepackage{dcolumn}
\usepackage{bm}
\usepackage{amsmath}
\usepackage{color}
\begin{document}

\title{Frictional weakening of vibrated granular flows}

\author{Abram H. Clark}
\affiliation{Department of Physics, Naval Postgraduate School, Monterey, California 93943, USA}
\author{Emily E. Brodsky}
\affiliation{Department of Earth and Planetary Sciences, University of California Santa Cruz, Santa Cruz, California 95064, USA}
\author{H. John Nasrin}
\affiliation{Naval Surface Warfare Center, Carderock Division, Bethesda, MD 20817, USA}
\author{Stephanie E. Taylor}
\affiliation{Department of Earth and Planetary Sciences, University of California Santa Cruz, Santa Cruz, tCalifornia 95064, USA}

\begin{abstract}

We computationally study the frictional properties of sheared granular media subjected to harmonic vibration applied at the boundary. Such vibrations are thought to play an important role in weakening flows, yet the independent effects of amplitude, frequency, and pressure on the process have remained unclear. Based on a dimensional analysis and DEM simulations, we show that, in addition to a previously proposed criterion for peak acceleration that leads to breaking of contacts, weakening requires the absolute amplitude squared of the displacement is sufficiently large relative to the confining pressure. The analysis provides a basis for predicting flows subjected to arbitrary external vibration and demonstrates that a previously unrecognized second process that is dependent on dissipation contributes to shear weakening under vibrations.

\end{abstract}

\date{\today}


\maketitle	
Recent years have seen dramatic advances in predictive constitutive laws for steady flows of dense granular media~\cite{daCruz2005,gdr2004dense,jop2006constitutive}, which are dominated by a Coulomb-like static friction coefficient $\mu_s$~\cite{bingham1917investigation,drucker1952soil}. Moreover, $\mu_s$ arises primarily from anisotropic, system-spanning contact networks~\cite{Peyneau2008,radjai2012fabric} that can be long-range correlated near the yield criterion~\cite{thompson2019}, leading to interesting nonlocal effects~\cite{bocquet2009kinetic,kamrin2012nonlocal,henann2013predictive} and avalanche-type behavior~\cite{dahmen2011simple}. The persistence of these mesoscale contact networks, often called ``force chains,'' during slow shear is predicated on the inherently dissipative nature of grain-grain interactions~\cite{jaeger96b,radjai1997force}, which arises from plasticity at individual contacts. 

Vibrations, which can be externally applied~\cite{melosh1979acoustic,melosh1996dynamical,Dijksman2011,taslagyan2015effect} or generated by the flow itself~\cite{vanderElst2012auto,taylor2017granular,degiuli2017friction,taylor2020reversible}, inject energy into the system, disrupting these contact networks and reducing $\mu_s$. Vibrations have been studied in granular pattern formation~\cite{melo1994transition,umbanhowar1996localized}, compaction~\citep{nowak98}, structural ordering~\cite{pica2007shear,picaciamarra2007tapping}, clogging~\cite{Caitano2021PRL}, and dense suspension rheology~\cite{Garat2022JoR}, but their impact on the resistance of shear flows has been underexplored. Seminal theoretical work \citep{melosh1979acoustic} and limited experiments \citep{Dijksman2011} have addressed parts of the problem, but the lack of a predictive framework for steady shear flows under vibration represents a significant gap in our understanding of a wide array of systems, including earthquakes, landslides, the results of impacts on asteroids, and the ability of the pharmaceutical industry to mass produce medication. 

In this letter, we use discrete-element method (DEM) simulations to systematically study the frictional properties of sheared, vibrated granular media. We vary amplitude and frequency of applied vibrations, as well as grain and other system properties. We find that previously proposed criteria based on contact breaking are insufficient to predict frictional weakening; the amplitude must also exceed a critical value that varies with pressure and grain-grain energy dissipation. Thus, in addition to contact breaking, the competition between vibration (energy injection) and dissipative grain-grain interactions plays a crucial role. We also find that frictional weakening stops when the frequency exceeds the elastic response frequency of the grains. Our results serve as the basis of a constitutive law that could be used to predict more complex steady flows subject to arbitrary external vibration.

\begin{figure}
    \raggedright (a) \hspace{30mm} (b) \\
    \centering
    \includegraphics[trim=0mm 28mm 0mm 20mm, clip, width=0.38\columnwidth]{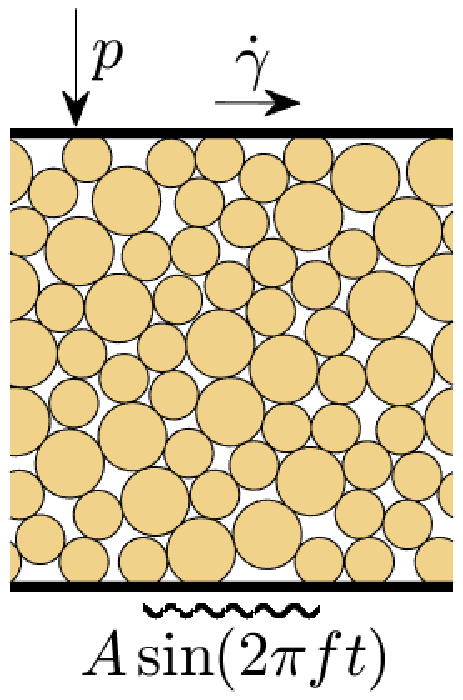}
    \centering
    \includegraphics[width=0.6\columnwidth]{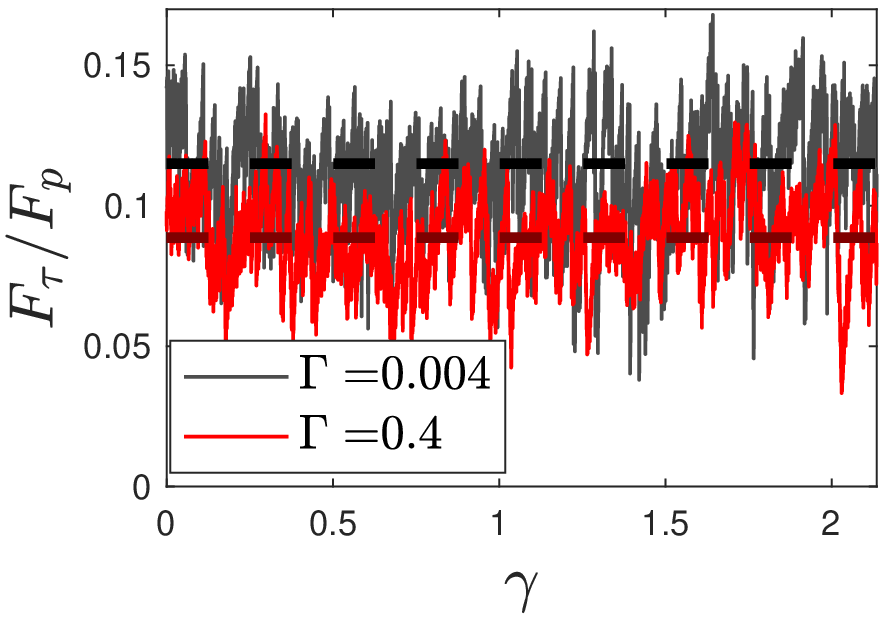} \\
    \raggedright (c) \\
    \centering
    \includegraphics[width=\columnwidth]{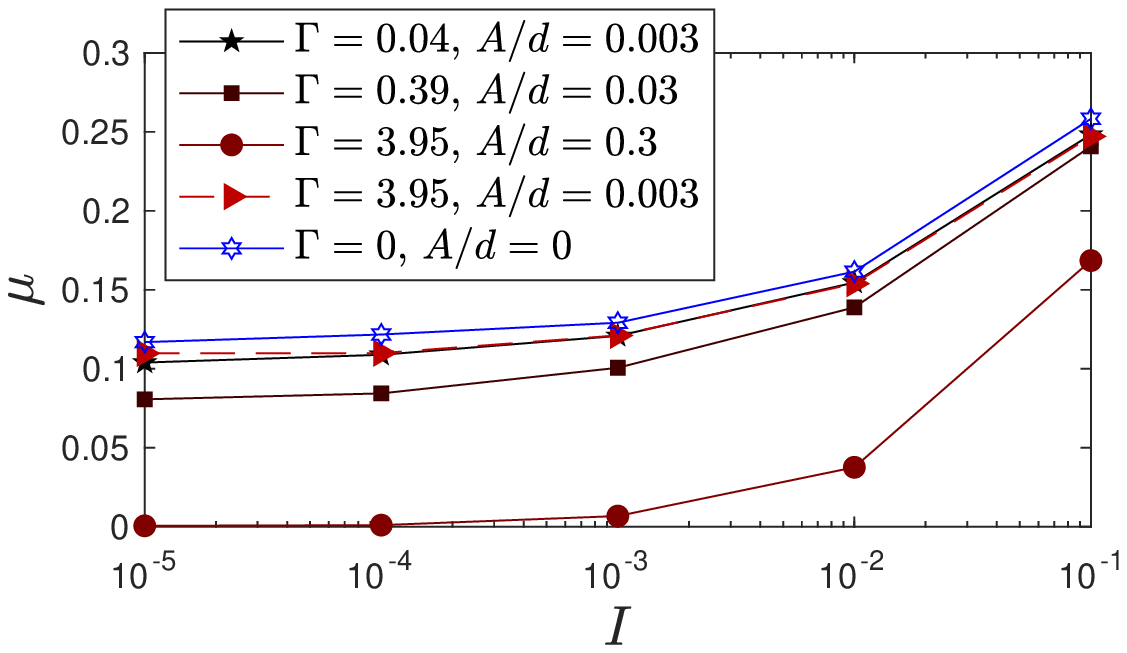}
    \raggedright (d) \\
    \centering
    \includegraphics[width=\columnwidth]{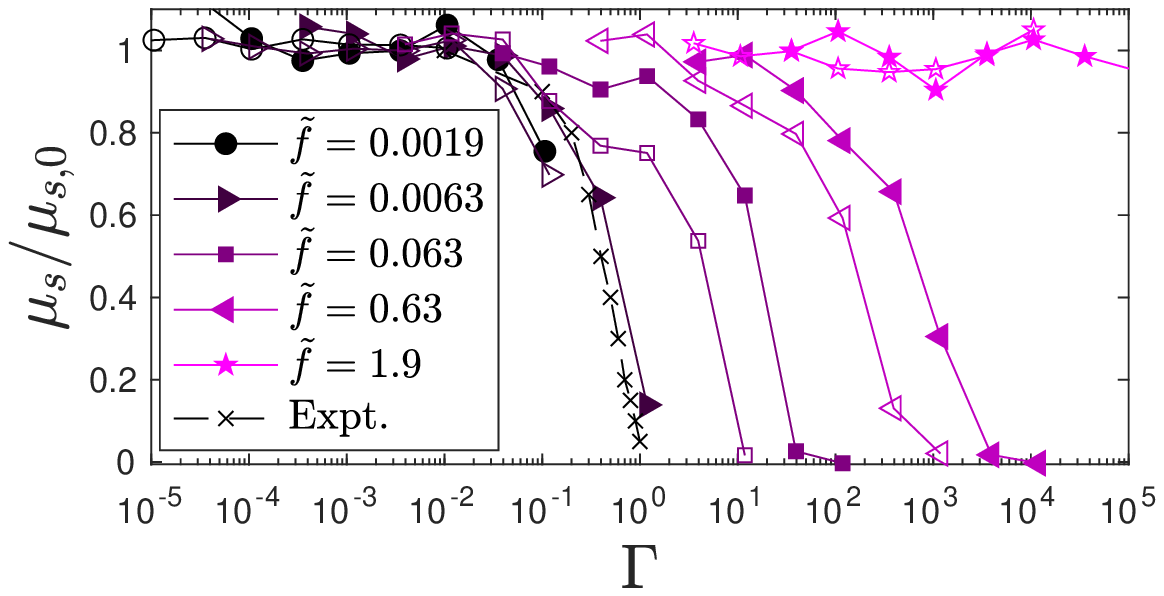}    
    \caption{(a) Depiction of the simulations; real simulations have rough walls. (b) Plots of $F_\tau/F_p$ versus shear strain $\gamma$ for two simulations with $I=10^{-5}$ after transients have subsided. The curves have $\Gamma = 0.004$, $\tilde{A} = 0.0003$ (dark gray) and $\Gamma = 0.4$, $\tilde{A} = 0.03$ (red). Dashed lines show the average, $\mu$. (c) $\mu$ versus $I$ for varying $\Gamma$ and $\tilde{A}$, with $e_n = 0.2$ and $\tilde{p}=10^{-3}$. (d) $\mu_s/\mu_{s,0}$ versus $\Gamma$ for varying $\tilde{f}$. Filled and open symbols correspond to $\tilde{p} = 10^{-5}$ and $\tilde{p} = 10^{-4}$, respectively; experimental data is from Fig. 2 of~\citet{Dijksman2011} using the smallest shear rate shown.}
    \label{fig:cartoon}
\end{figure}

The fundamental question we consider is: when do vibrations of amplitude $A$ and frequency $f$ cause frictional weakening, i.e., $\mu_s$ to decrease significantly? We begin with a dimensional analysis of simple shear of a granular system subjected to such vibration, as in Fig.~\ref{fig:cartoon}(a). The shear rate $\dot{\gamma}$ is imposed by moving the top wall, and the normal stress $p$ is imposed by applying a fixed external force per area to the top wall. $\tau$ is the average force per area required to maintain $\dot{\gamma}$. Grains properties include diameter $d$, mass density $\rho$, elastic modulus $E$, the restitution coefficient $e_n$, and surface friction coefficient $\mu_g$, and possibly others (e.g., shape). Neglecting vibration, five dimensionless groups are necessary to characterize such a system: the material friction coefficient $\mu = \tau/p$, inertial number $I = \dot{\gamma}d\sqrt{\rho/p}$, dimensionless pressure $\tilde{p} = p/E$, $e_n$, and $\mu_g$. If $\tilde{p}$ is small enough to be irrelevant, then, for fixed $\mu_g$ and $e_n$, we can write $\mu(I)$, which can often be well approximated by $\mu(I) = \mu_s + bI^a$, where $\mu_s$ is the static friction coefficient \citep{jop2006constitutive, daCruz2005} and $a$ varies with $\mu_g$~\cite{FavierdeCoulomb2017rheology,srivastava2021viscometric}. 

Including $A$ and $f$ requires two more dimensionless numbers. We choose $\tilde{A} = A/d$ and $\Gamma = {A(2\pi f)^2 \rho d}/{p}$, which is the ratio of $A(2 \pi f)^2$, the peak acceleration from the vibration, to $p/\rho d$, the acceleration resulting from the applied normal stress. Experiments by \citet{Dijksman2011} on vibration of a sheared granular bed with a free surface (using the gravitational acceleration $g$ instead $p/\rho d$) found that $\mu_s \approx 0$ when $\Gamma>1$, corresponding to when the vibrated bottom wall will lose contact with the particles, allowing them to rearrange. $\Gamma$ has also been used in a variety of other systems~\cite{umbanhowar1996localized,Caitano2021PRL,Garat2022JoR}. At high $f$, the dimensionless number $\tilde{f} = (\Gamma/\tilde{A}\tilde{p})^{1/2} = 2\pi f d\sqrt{\rho/E}$ becomes relevant as the ratio of $f$ to the elastic frequency of grains.

The classic theory of~\citet{melosh1979acoustic}, which has been heavily utilized in the geophysical sciences \citep{collins2003, Johnson2016}, proposed that fluidization occurs when the peak acoustic pressure $s$  exceeds the confining pressure $p$, i.e., $s/p>1$, breaking grain-grain contacts.  In an elastic wave, peak pressure can be written as $s = \rho c \omega A$~\cite{halliday1992physics}, where $c = \sqrt{E/\rho}$ is a wave speed and $\omega = 2\pi f$. Thus, the condition $s/p>1$ in terms of the parameters of this paper is  $(\tilde{A} \Gamma/\tilde{p})^{1/2}>1$. Like the $\Gamma$ framework, acoustic fluidization uses a \textit{single criterion related to contact breaking}. Neither framework is set up to disentangle the independent effects of $A$ and $f$ and thus neither can uncover other criteria or processes. Other work has noted that additional parameters are likely necessary in other vibrated flows \citep{Caitano2021PRL}, but none have clarified what the correct approach might be for the geologically important situation of shear flows. 


We implement a vibrating shear flow using DEM simulations using LAMMPS~\cite{LAMMPS}. These simulations involve simple shear of an assembly of $N$ spherical grains via the motion of a top wall with imposed vibrations at the bottom wall, as depicted in 2D in Fig.~\ref{fig:cartoon}(a). The horizontal dimensions are both periodic with length $L$. Our results are insensitive to the system size and aspect ratio, which we verify by changing $L$ and $N$ as illustrated in Supplemental Material~\cite{suppmat}. This means our results are not primarily due to vibrational resonance based on $L$ or the ability of phonons to propagate across the system.

Grain-grain forces consist of a normal repulsive term, characterized by spring constant $k_n = Ed$, and a viscoelastic damping force for normal contacts, characterized by damping coefficient $\gamma_n$ that is related to a normal restitution coefficient $e_n$~\cite{shafer1996force,suppmat}. We focus on frictionless particles in the main text. In the Supplemental Material~\cite{suppmat}, we include grain-grain friction via the Cundall-Strack~\cite{cundall79} approach, as well as 2D simulations with bumpy particles~\cite{papanikolaou2013isostaticity,clark2017role}. We also show data for Hertzian contacts.  In this paper we focus  on the robust results that are  qualitatively similar results for all cases, regardless of spatial dimension, friction, force law, or grain shape. Grain diameters are normally distributed with mean $d$ and standard deviation of $0.2d$. Top and bottom walls are rough, created via rigid assemblies of the same particles used in the flow, to ensure a no-slip boundary between the wall and the granular assembly. The wall-grain forces are computed as the sum of forces between wall particles and particles in the flow. We approximate $\mu$ via forces on the walls, which neglects second-order effects related to normal stress difference; these are very small, especially for frictionless particles at low inertial number, as shown by~\citet{srivastava2021viscometric}.

We impose a confining (downward) force $F_p = p L^2$ on the top wall as well as a horizontal velocity $v$; motion of the wall in the third dimension is not allowed. We measure the total horizontal force $F_\tau = \tau L^2$ on the wall due to all wall-grain contacts. After initial transients have decayed, $F_\tau$ fluctuates around a constant value, as shown in Fig.~\ref{fig:cartoon}(b). For each simulation, we measure $\mu = \langle F_\tau \rangle / F_p = \tau/p$ as the average, steady-state friction coefficient. The height $H$ fluctuates around an average value $\langle H \rangle \sim Nd^3/L^2$, and we measure the strain rate $\dot{\gamma} = v/\langle H\rangle$ and thereby the inertial number $I = \dot{\gamma} d \sqrt{\rho/p}$. We do not allow the lower wall to move except for an imposed vertical harmonic displacement with amplitude $A$ and frequency $f$.

The output of each simulation is $\mu$ as a function of $I$, $\tilde{p}$, $e_n$, $\Gamma$, and $\tilde{A}$. Figure~\ref{fig:cartoon}(b) shows results from two typical simulations with differing $\Gamma$. As expected, $\mu$ is lower for larger $\Gamma$. We observe very little dilation for all results we show here, i.e., $\langle H \rangle$ does not vary strongly with $\tilde{A}$ or $\Gamma$. The time step is 100 times smaller than the time scale for a grain-grain collision (see Supplemental Material~\cite{suppmat}), which is sufficient to resolve vibration frequencies for $\tilde{f} < 10$. We also verify that the vibrations imposed on the bottom traverse the system by measuring their perturbation on the top wall (Figure~\ref{fig:cartoon}(b)). Because of the large shear and correspondingly large number of samples in the mean values reported for each simulation, uncertainty estimates based on bootstrap resampling \citep{efron1982} are between 0.1 and 0.5\%. This uncertainty is smaller than the symbols here and in the remainder of the figures.

Figure~\ref{fig:cartoon}(c) investigates the shear rate-dependent friction by measuring $\mu(I)$ curves with $e_n = 0.2$, $\tilde{p} = 10^{-4}$, and varied $\Gamma$ and $\tilde{A}$ (including $\Gamma = \tilde{A} = 0$). For each curve, $\mu$ is roughly constant for $I\leq 10^{-4}$, corresponding to $\mu = \mu_s$. All measurements of $\mu_s$ use $I = 10^{-4}$, and we verify with selected simulations at $I = 10^{-5}$ that we are in the slow-shear limit. With $\Gamma = \tilde{A} = 0$, $\mu_s \approx 0.12$, as expected for frictionless spheres~\cite{Peyneau2008,thompson2019}. For the remainder of the paper, we denote $\mu_{s,0}$ as the friction coefficient in the limit of low inertial number and no applied vibration (e.g., for stiff, frictionless spheres, $\mu_{s,0} \approx 0.12$). 

The results in Fig.~\ref{fig:cartoon}(c) demonstrate that frictional weakening cannot be predicted from $\Gamma$ alone.  
For three curves, we keep constant frequency and increase amplitude, and $\mu$ decreases as $\Gamma$ and $\tilde{A}$ increase, as expected. However, an additional $\mu(I)$-curve with the largest value  of $\Gamma$ but  a higher $f$ and lower $\tilde{A}$ results in friction similar to the low-$\Gamma$ result.

Additional support for the need for multiple parameters, and thus multiple mechanisms, to describe vibrational weakening comes from Fig.~\ref{fig:cartoon}(d), which shows the normalized friction $\mu_s/\mu_{s,0}$ as a function of $\Gamma$ with $f$ held constant and $\tilde{A}$ increased from $10^{-4}$ to 0.3. The low-$\tilde{f}$ simulations are an excellent match to the experimental data from~\citet{Dijksman2011}. However, as $f$ and $\tilde{p}$ are varied, the value of $\Gamma$ where $\mu_s$ transitions to zero varies dramatically, over more than three orders of magnitude. As in Figure~\ref{fig:cartoon}(c) variations cannot be mapped simply as a function of $\Gamma$. For $\tilde{f}>1$, no weakening occurs since $f$ exceeds the elastic frequency of the grains. The significant dependence on $\tilde{A}$, $\tilde{p}$ and $\tilde{f}$ requires a more complete description of both the parameters and the physics.

\begin{figure}
\raggedright (a) \\
    \centering
    \includegraphics[width=\columnwidth]{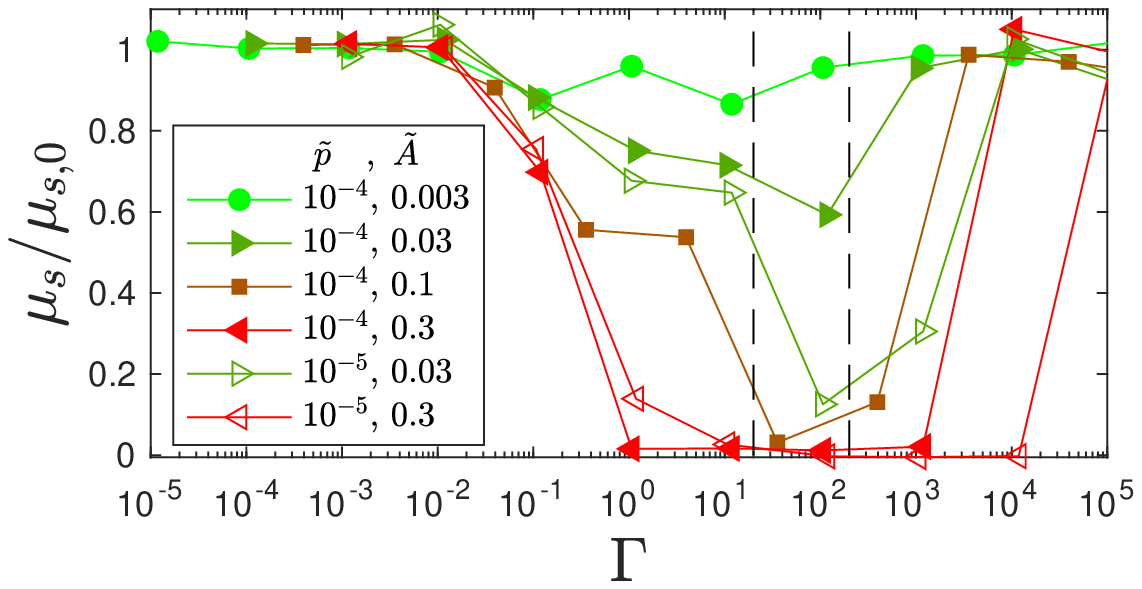}
\\ \raggedright (b) \\ \centering
    \includegraphics[width=\columnwidth]{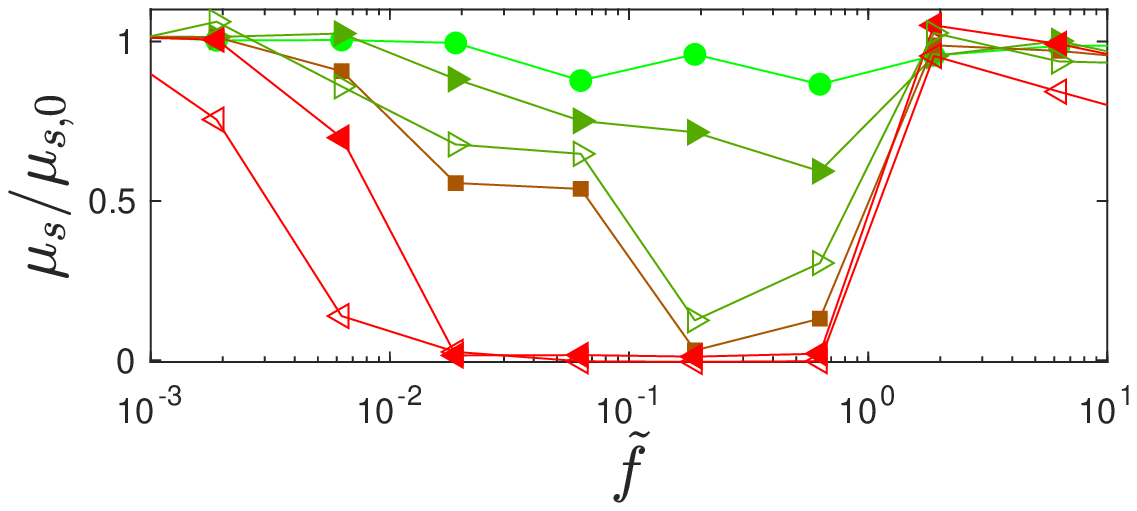}
    \caption{(a) $\mu_s/\mu_{s,0}$ versus $\Gamma$, where $f$ is varied and $\tilde{A}$ is held constant. We use the data between the vertical dashed lines to characterize the dependence of $\mu_s$ on $\tilde{A}$ and $\tilde{p}$; see text for discussion. (b) $\mu_s/\mu_{s,0}$ versus $\tilde{f}$ for the same data shown in (a), showing that $\tilde{f}>1$ corresponds to the rise of $\mu_s$ at high $\Gamma$.}
    \label{fig:mu_correct}
\end{figure}

When we vary $\Gamma$ by varying $f$ and holding $A$ fixed, a clearer picture emerges, as shown in Fig.~\ref{fig:mu_correct}. Figure~\ref{fig:mu_correct}(a) again shows the normalized friction $\mu_s/\mu_{s,0}$ as a function of $\Gamma$ for sheared, vibrated, frictionless spheres with $e_n = 0.2$, but with each curve having a fixed $\tilde{A}$ and only the frequency $f$ varied. For $\Gamma < 0.1$, all curves have $\mu_s \approx \mu_{s,0}$. For $\Gamma>0.1$, we find $\mu_s$ begins decreasing in a way that depends on $\tilde{A}$ and $\tilde{p}$. Thus, $\Gamma<0.1$ always corresponds to no frictional weakening. We also observe no frictional weakening at very high $\Gamma$; Fig.~\ref{fig:mu_correct}(b) demonstrates that this is due to $\tilde{f}>1$

However, $\Gamma>1$ and $\tilde{f}<1$ are still not sufficient to predict frictional weakening; $\tilde{A}$ must be also large enough. This indicates an additional process at play. Perhaps the amplitude needs to be large enough to induce sufficient rearrangements to disrupt the force network. Weakening will not occur if these amplitudes are not high enough for a given $\tilde{p}$, regardless of $\Gamma$. We now consider how $\mu_{\rm min}$ varies with $\tilde{A}$ and $\tilde{p}$ at fixed $\Gamma$.



\begin{figure}
    \raggedright (a) \\
    \centering
    \includegraphics[trim=2mm 2mm 5mm 5mm,clip,width=\columnwidth]{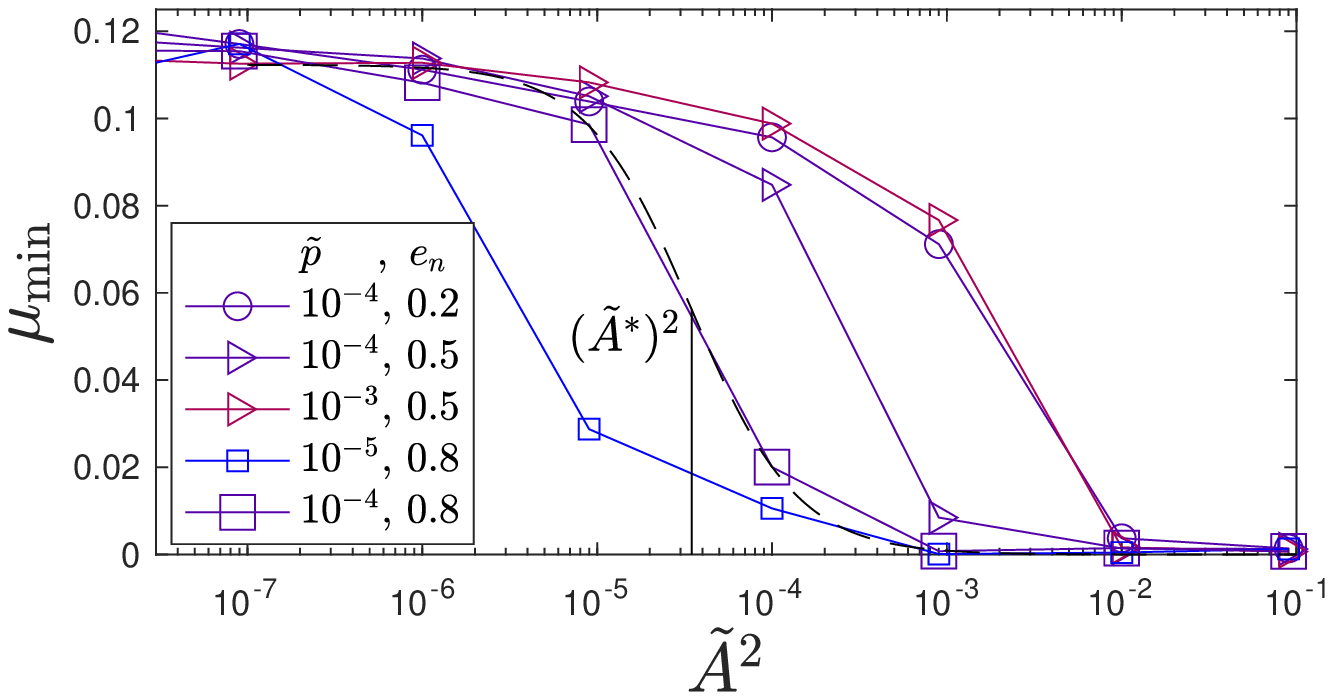} \\
    \raggedright (b) \hspace{40mm} (c) \\
    \includegraphics[width=0.49\columnwidth]{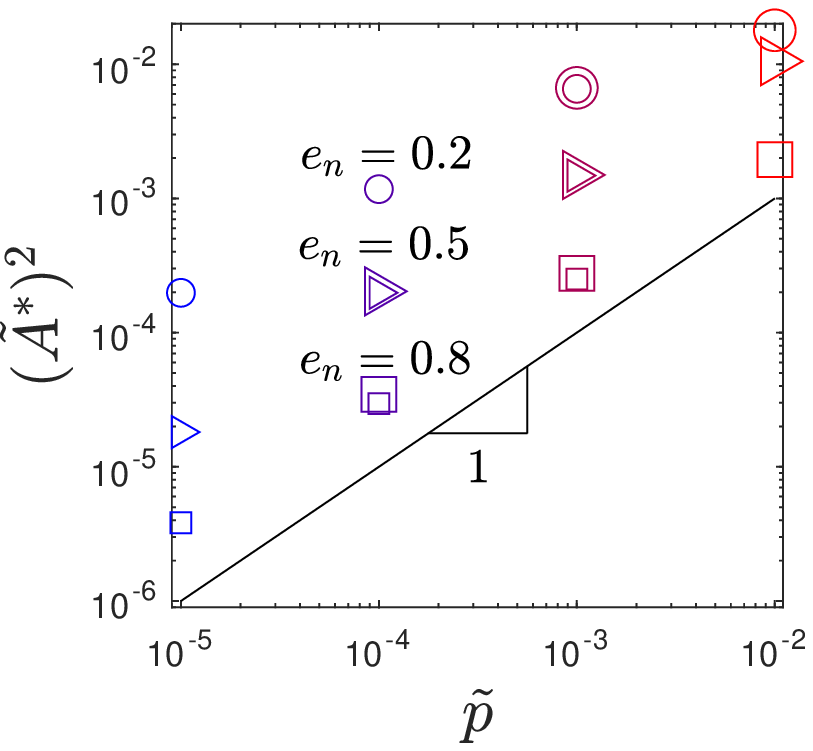}
    \includegraphics[width=0.49\columnwidth]{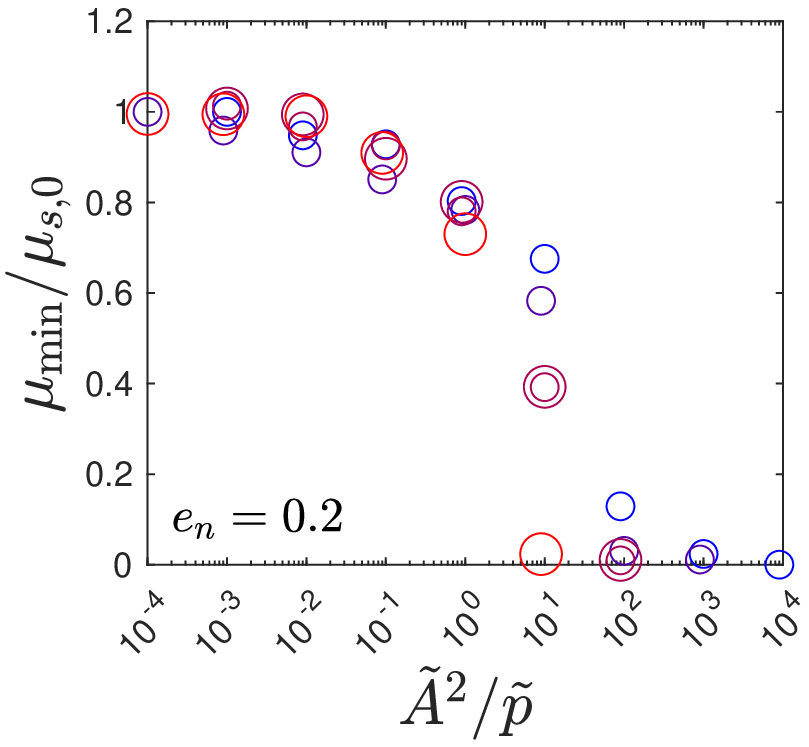} \\
    \raggedright (d) \hspace{40mm} (e) \\
    \includegraphics[width=0.49\columnwidth]{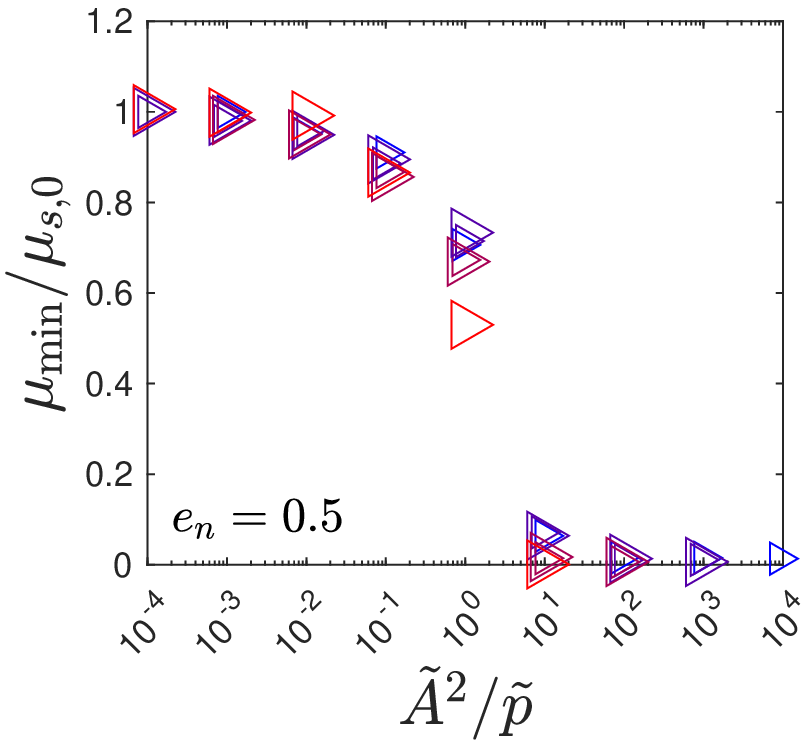}
    \includegraphics[width=0.49\columnwidth]{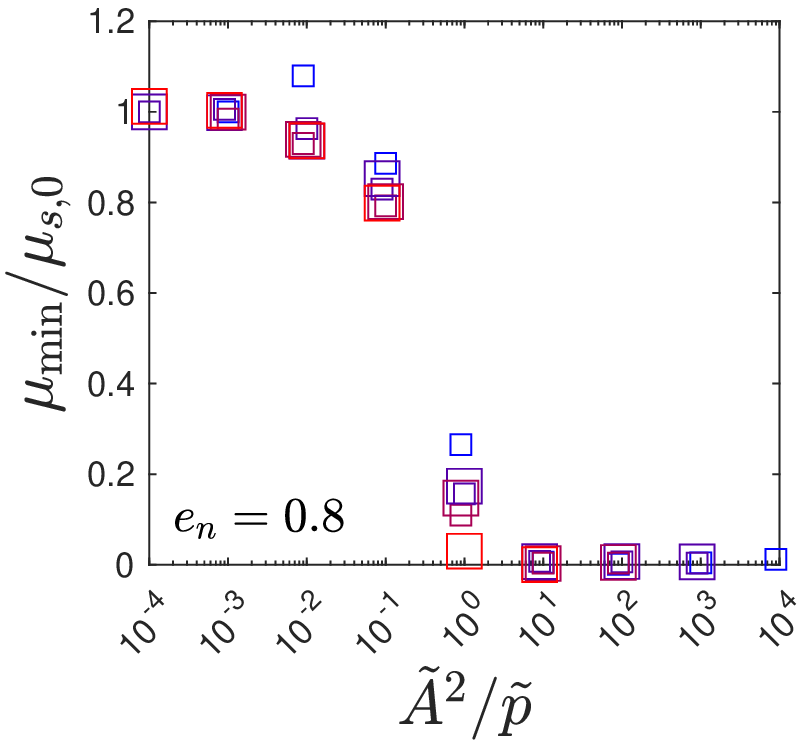} \\
    \caption{(a) $\mu_{\rm min}$, measured between the dashed lines in Fig.~\ref{fig:mu_correct}(a), is plotted as a function of $\tilde{A}^2$ for different $\tilde{p}$ and $e_n$. Dashed black line shows a fit to a sigmoid-like function, $\mu_{\rm min} = \frac{\mu_{s,0}}{2}\{1-\tanh[\log_{10}(\tilde{A}/\tilde{A}^*)^2]\}$, to the data for $\tilde{p} = 10^{-4}$ and $e_n = 0.8$. The fit estimates $\tilde{A}^*$ as the value of $\tilde{A}$ where $\mu_s/\mu_{s,0} = 1/2$, which we use as the characteristic value of $\tilde{A}$ for frictional weakening. (b) $(\tilde{A}^*)^2$ versus $\tilde{p}$ for $e_n = 0.2$ (circles), $e_n = 0.5$ (triangles), and $e_n = 0.8$ (squares). Color denotes $\tilde{p}$; large and small symbols have different values of $E$, confirming that $\tilde{p}$ captures the scaling. All data are approximately captured by $(\tilde{A}^*)^2 \propto \tilde{p}$. (c-e) $\mu_{\rm min}/\mu_{s,0}$ versus $\tilde{A}^2/\tilde{p}$ for (c) $e_n = 0.2$, (d) $e_n = 0.5$, and (e) $e_n = 0.8$, with the same symbol convension as in (b).}
    \label{fig:mu_min}
\end{figure}

We measure $\mu_{\rm min}$ as lowest value of $\mu(\Gamma)$ between the dashed lines shown in Fig.~\ref{fig:mu_correct}(a), i.e., $\Gamma \approx 100$. This definition is selected so as to keep $\Gamma$ fixed throughout the comparison. We repeat all simulations for $e_n=0.5$ and $e_n = 0.8$ and find very similar results to those shown in Fig.~\ref{fig:mu_correct}. Figure~\ref{fig:mu_min}(a) shows curves of $\mu_{\rm min}$ versus $\tilde{A}$ for different combinations of $e_n$ and $\tilde{p}$. For small $\tilde{A}$, $\mu_{\rm min} \approx \mu_{s,0}$, and $\mu_{\rm min}$ decreases from $\mu_{s,0}$ to 0 at a characteristic value of $\tilde{A}$, denoted $\tilde{A}^*$, that depends on $\tilde{p}$ and $e_n$. 

We estimate $\tilde{A}^*$ by fitting a sigmoid-like curve to the data in Fig.~\ref{fig:mu_min}(a) to extract $\tilde{A}^*$ as the value where $\mu_{\rm min} = \mu_{s,0}/2$. Figure~\ref{fig:mu_min}(b) shows that $(\tilde{A}^*)^2 \propto \tilde{p}^\beta$. Best fits give $\beta$ near 1 for all three values of $e_n$: $\beta = 0.90\pm 0.02$, $0.92\pm 0.13$, and $0.76\pm 0.09$ for $e_n = 0.8$, 0.5, and 0.2, respectively, where the data point with $\tilde{p} = 10^{-2}$ is disregarded for $e_n = 0.2$. We assume $\beta \approx 1$, and the fact that $\beta<1$ may be due to additional contacts leading to more dissipation at higher pressure. This is consistent with the deviation at $e_n = 0.2$ especially for the highest pressure; future analysis may provide some further insight. Decreasing $e_n$ corresponds to higher $(\tilde{A}^*)^2$ at fixed $\tilde{p}$, meaning more vibration amplitude is required at higher dissipation rates for frictional weakening to occur. Figure~\ref{fig:mu_min}(c), (d), and (e) show $\mu_{\rm min}/\mu_{s,0}$ as a function of $\tilde{A}^2/\tilde{p}$ for all three values of $e_n$. These plots show a reasonable data collapse with $\mu_{\rm min}$ decreasing to 0 for $\tilde{A}^2/\tilde{p}$ of approximately $10^2$ for $e_n = 0.2$, $10^1$ for $e_n = 0.5$, and $10^0$ for $e_n = 0.8$. This significant variation with $e_n$ highlights the crucial role of grain-grain dissipation.

\begin{figure}
    \centering
    \includegraphics[width=\columnwidth]{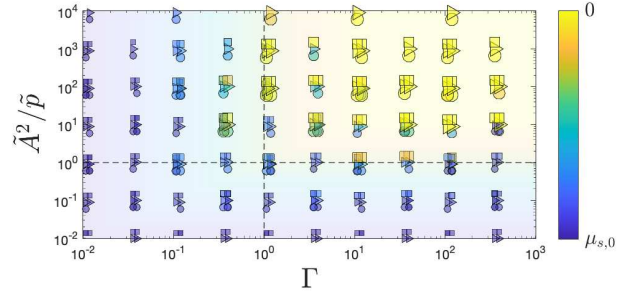}
    \caption{A phase diagram of $\mu_s$ as a function of $\Gamma$ and $\tilde{A}^2/\tilde{p}$. Symbol shapes correspond to different $e_n$ and are slightly shifted for visibility, with circles for $e_n = 0.2$ (shifted down), triangles for $e_n = 0.5$, and squares for $e_n = 0.8$ (shifted up). Symbol size and color represents the amount of frictional weakening. Dashed lines show $\Gamma = 1$ and $\tilde{A}^2/\tilde{p} = 1$.}
    \label{fig:phase_diag}
\end{figure}

Our results can be summarized in the phase diagram shown in Fig.~\ref{fig:phase_diag}.  As in prior work, when $\Gamma>1$ contacts can be broken, but the current simulations show that large $\Gamma$ corresponds to frictional weakening only when $\tilde{A}^2/\tilde{p}$ is large. The magnitude of $\tilde{A}^2/\tilde{p}$ required depends on $e_n$, since more amplitude at the boundary is required to give individual grains sufficient vibrational energy to disrupt the contact network and reduce $\mu_s$. This dependence on $e_n$ can be seen in Fig.~\ref{fig:phase_diag} at, e.g., $\Gamma = 10$, $\tilde{A}^2/\tilde{p}=1$. The controlling parameter $\tilde{A}^2/\tilde{p}$ might be interpreted as the ratio of a force scale related to the amplitude of the vibration coupled to the particle stiffness, $E A^2$, to a characteristic force on a particle due to the confining pressure, $p d^2$. Future work may shed further light on the interpretation of this criterion.

Importantly, the phase diagram shows that prior work based on a single criterion could overpredict fluidization in geologically relevant situations. For instance, in the nearfield of an impact or shallow fault zone, a wave with a frequency of  10~Hz and amplitude 1~mm can interact with sand-sized particles with with $d = 0.1$~mm and $E = 70$~GPa at pressure 0.25~MPa, corresponding to 10~m depth. These reasonable values correspond to  $\Gamma \approx 4\times 10^{-6}$ and $\tilde{A}^2/\tilde{p} \approx 3 \times 10^{7}$. This is far in the upper left quadrant of Fig.~\ref{fig:phase_diag}, where no fluidization would occur. However, the fluidization condition of~\citet{melosh1979acoustic} would predict fluidization, since $(\tilde{A} \Gamma/\tilde{p})^{1/2} \approx 3$, which is greater than the threshold of 1. Practical applications of acoustic fluidization theory to observations have adjusted the acoustic wavelength (and hence frequency) to match observations where independent constraints are not possible \citep{riller2018}. Our results may be useful in reconsidering such inferences, as well as in other situations where weakening is experimentally observed due to acoustic excitation~\cite{vanderElst2012auto,lu2007}. 


In summary, we find that frictional weakening requires both sufficiently high acceleration and amplitude, appropriately normalized. The acceleration criterion ($\Gamma$) can be attributed to a need to break individual contacts as noted by many prior works \citep{melosh1979acoustic, Dijksman2011}. The amplitude criterion ($\tilde{A}^2/\tilde{p}$) shows the need for an additional process that is sensitive to the degree of dissipation, which provides an important clue. Dissipation is required to maintain the mesoscale network structures or ``force chains'' during shear~\citep{jaeger96b,radjai1997force,wolf1998dissipation}. These structures are known to control the macroscopic frictional properties of granular media~\cite{Peyneau2008,Radjai2014PRL}. Thus we speculate that the amplitude criterion relates to the disruption of these structures. Simple contact breaking is not sufficient if the latent mesoscale structure is preserved; sufficiently large amplitude is required to break them up. The lack of inclusion of this amplitude criterion results in an overprediction of frictional weakening. More importantly, the recognition of an additional process positions the field to investigate the correct criteria to determine the efficacy of frictional weakening in some of nature's most important granular flows.

\begin{acknowledgments}
We gratefully acknowledge funding from Army Research Office under grants W911NF1510012 and W911NF2220044 and the Office of Naval Research under grant  N0001419WX01519. We also thank Jeffrey Haferman and Bruce Chiarelli for help with high-performance computing at NPS and the Ship Engineering and Analysis Technology Center for help with high-performance computing at NSWC Carderock.
\end{acknowledgments}

\bibliography{references}

\end{document}